\documentclass[runningheads]{llncs}
\usepackage[T1]{fontenc}
\usepackage{graphicx}
\usepackage{amsmath,amsfonts}
\usepackage{algorithmic}
\usepackage{algorithm}
\usepackage{array}
\usepackage[caption=false,font=normalsize,labelfont=sf,textfont=sf]{subfig}
\usepackage{textcomp}
\usepackage{stfloats}
\usepackage{url}
\usepackage{verbatim}
\usepackage{graphicx}
\usepackage{cite}
\usepackage{multirow}
\usepackage{flushend}
\usepackage{caption}
\usepackage{subcaption}
\usepackage{comment}

\usepackage{diagbox}

\usepackage[table]{xcolor}

\usepackage{color}
\begin{document}
\title{AutoML in Cybersecurity: An Empirical Study}
\author{Sherif Saad\inst{1}\and
Kevin Shi\inst{1}\and
Mohammed Mamun\inst{2} \and
Hythem Elmiligi\inst{3}}
\authorrunning{S. Saad et al.}
\institute{School of Computer Science University of Windsor, ON, Canada
\email{\{shsaad,shi12z\}@uwindsor.ca}\and
National Research Council, NB, Canada\\
\email{mohammad.mamun@nrc-cnrc.gc.ca}\and
University of Victoria, BC, Canada\\
\email{haytham@ieee.org}}
\maketitle              % typeset the header of the contribution
\begin{abstract}
Automated machine learning (AutoML) has emerged as a promising paradigm for automating machine learning (ML) pipeline design, broadening AI adoption. Yet its reliability in complex domains such as cybersecurity remains underexplored. This paper systematically evaluates eight open-source AutoML frameworks across 11 publicly available cybersecurity datasets, spanning intrusion detection, malware classification, phishing, fraud detection, and spam filtering. Results show substantial performance variability across tools and datasets, with no single solution consistently superior. A paradigm shift is observed: the challenge has moved from selecting individual ML models to identifying the most suitable AutoML framework, complicated by differences in runtime efficiency, automation capabilities, and supported features. AutoML tools frequently favor tree-based models, which perform well but risk overfitting and limit interpretability. Key challenges identified include adversarial vulnerability, model drift, and inadequate feature engineering. We conclude with best practices and research directions to strengthen robustness, interpretability, and trust in AutoML for high-stakes cybersecurity applications.
\end{abstract}


\section{Introduction}
AutoML encompasses techniques and processes that automate key stages of the ML workflow, making model development more accessible to non-experts. Its primary goal is to reduce the manual effort required in designing and training ML models, thereby enabling individuals with limited expertise to leverage machine learning capabilities~\cite{10.1145/3544548.3581082}. Typical ML workflows involve tasks such as data preprocessing, feature engineering, model selection, hyperparameter tuning, and evaluation. AutoML frameworks streamline these steps by automatically searching for optimal model architectures and hyperparameters to achieve strong performance on a given dataset~\cite{9978249}.

Despite the availability of numerous AutoML tools, systematic evaluations of state-of-the-art solutions remain limited. It is still unclear how automatically generated pipelines compare to models crafted by experts, particularly in complex, high-stakes domains such as cybersecurity. The reliability and effectiveness of AutoML in real-world cybersecurity applications therefore remain uncertain.

Performance metrics for the same AutoML tool can vary substantially due to multiple factors. Dataset characteristics—such as distribution, complexity, and noise—strongly influence outcomes. Performance also depends on feature engineering strategies, algorithm selection, hyperparameter settings, evaluation metrics, cross-validation methods, and preprocessing techniques. Randomness in weight initialization, data shuffling, or variations across software versions, libraries, and hardware further contribute to variability. While prior work has benchmarked AutoML tools on general ML tasks, this study provides, to our knowledge, the first large-scale, systematic evaluation of AutoML tools specifically for cybersecurity. This domain presents unique challenges—including adversarial threats, data imbalance, concept drift, and the need for interpretable models—that are not addressed in conventional AutoML studies. Our contributions are as follows:

\begin{enumerate}
    \item We present the first comprehensive empirical evaluation of eight widely used AutoML tools on 11 publicly available cybersecurity datasets, spanning intrusion detection, malware detection, phishing, fraud detection, and spam filtering.
    \item We evaluate performance not only in terms of accuracy, but also balanced accuracy, runtime efficiency, and model complexity, offering a multidimensional benchmark to guide practitioners in real-world applications.
    \item We identify persistent challenges—such as over-reliance on tree-based models, vulnerability to data leakage, and the lack of mechanisms for adversarial robustness or model drift—and propose best practices to address these gaps.
\end{enumerate}

By addressing these issues, our work advances the understanding of AutoML’s readiness for adversarial domains such as cybersecurity and provides practical guidance for researchers and practitioners. In particular, we aim to support software engineers and cybersecurity developers considering AutoML by helping establish trust and assess tool reliability.

The remainder of the paper is organized as follows. Section~\ref{sec:review} reviews related studies on AutoML evaluation. Section~\ref{sec:automl4cyber} discusses the challenges of applying AutoML in cybersecurity. Section~\ref{sec:method} describes our methodology. Section~\ref{sec:experiment} presents the experimental results and analysis. Finally, Section~\ref{sec:conclusion} concludes the paper and outlines future directions.

\section{Literature Review}
\label{sec:review}

Several studies have evaluated state-of-the-art AutoML tools for model selection. Tuggener et al.~\cite{tuggener2019automated} compared four tools—Auto-Sklearn, TPOT, Data Science Machine (DSM), and Portfolio Hyperband—with DSM serving as a baseline. Auto-Sklearn, TPOT, and DSM were tested on regression, binary classification, and multi-class classification, while Portfolio Hyperband was limited to binary classification. The results showed that Auto-Sklearn, TPOT, and DSM consistently outperformed the DSM baseline, though DSM remained competitive when restricted to a small set of proven options.

Ge et al.~\cite{ge2020analysis} evaluated Auto-Sklearn, TPOT, AutoKeras, and AutoGluon, finding AutoGluon to be the most effective overall. The authors concluded that all four tools produced robust, accurate models with a shared objective of being user-friendly. Similarly, Truong et al.~\cite{truong2019towards} compared commercial and open-source AutoML tools, noting differences in model selection and hyperparameter optimization approaches. Commercial tools such as H2O-Driverless AI, DataRobot, and Darwin automatically detected data schemas, performed feature engineering, and supported model interpretation, while open-source tools focused primarily on model training and selection. Their results showed H2O AutoML performing best in binary classification and regression, while AutoKeras achieved superior results in multi-class classification. The authors recommended AutoKeras and H2O AutoML for production environments requiring speed and performance stability.

Beyond performance, user trust and transparency are also critical. Drozdal et al.~\cite{drozdal2020trust} emphasized that increasing transparency improves user confidence in AutoML tools. Similarly, Xin et al.~\cite{xin2021whither}, in a study involving 16 participants, argued that AutoML should not replace human involvement but instead support workflows. Nagarajah et al.~\cite{nagarajah2019review} highlighted that current systems cannot serve as universal solutions, and that hyperparameter tuning benefits significantly from ensembling and meta-learning.

Overall, the literature shows that most efforts in AutoML research focus on model selection and hyperparameter optimization. However, there is limited work directly comparing AutoML-selected models with those manually tuned by domain experts. Moreover, commonly used datasets in evaluations are often either too generic or overly simplistic, lacking application to a concentrated problem domain. This is particularly evident in cybersecurity, where only a handful of studies have assessed AutoML tools.

One of the earliest cybersecurity-related evaluations involved SMS spam filtering. Saeed et al.~\cite{saeed2021comparison} compared MLJAR-supervised, H2O, and TPOT, reporting that H2O AutoML with stacked ensembles achieved the best results in log loss, true positive, and true negative rates. Similarly, Suleiman et al.~\cite{suleiman2020deep} used H2O AutoML to evaluate deep learning, random forest, and Naïve Bayes for SMS spam filtering, finding random forest to be the best-performing algorithm. Purwanto et al.~\cite{purwanto2021man} compared six AutoML tools on ten phishing datasets and assessed whether AutoML models outperformed handcrafted models. They found that AutoML models often matched or exceeded expert models, but also noted the tools’ reliance on supervised learning, reinforcing the need for domain expertise in phishing detection.

Recent studies have further explored AutoML in cybersecurity. Opara et al.~\cite{opara2022auto} applied AutoML techniques to botnet detection, reporting promising results for automated model selection. Glavan et al.~\cite{glavan2023autoencoders} emphasized the importance of domain-specific feature engineering within AutoML pipelines for cybersecurity tasks. Alrowais et al.~\cite{alrowais2023automated} proposed broader frameworks to integrate AutoML into cybersecurity workflows for threat detection and mitigation. While these works target specific problems or propose frameworks, to the best of our knowledge, our study is the first to provide large-scale, empirical benchmarking of AutoML tools across diverse cybersecurity applications—including malware detection, phishing, fraud detection, and intrusion detection—with direct comparisons to handcrafted models.

\section{A\MakeLowercase{uto}ML Challenges in Cybersecurity}
\label{sec:automl4cyber}

Cybersecurity threats can be modeled as data-driven problems and addressed with machine learning. However, applying ML to cybersecurity introduces unique challenges that make it more difficult than in other domains. These challenges, well recognized in prior studies~\cite{Amit2018MachineLI, ceschin2020machine}, stem from the complexity and adversarial nature of cybersecurity environments. In this section, we focus on the challenges most relevant to AutoML, as identified from prior literature and our evaluation of selected tools.

\textbf{Feature Engineering and Extraction:}
Cybersecurity data is highly diverse and often requires domain-specific feature engineering. Features rarely conform to standard numerical, categorical, or nominal types. For example, port numbers are numeric values but cannot be treated as continuous variables; similarly, binary file or protocol names (e.g., HTTP, SMTP, SSH) carry semantic meaning not captured by simple categorical encodings. Conventional methods such as one-hot encoding therefore fail to represent their inherent structure.

To address this, AutoML tools should incorporate automated or semi-automated feature engineering methods. Features curated and validated by domain experts in the literature could be organized into catalogs or feature stores and then operationalized as reusable design patterns. Automating such domain-specific feature engineering would significantly enhance AutoML’s applicability in cybersecurity. We adopted this approach while preparing several datasets in our study. However, while automated feature engineering is important for AutoML, it may introduce other challenges related to model explainability and interoperability when utilizing AutoML \cite{bosch2021automl}.

\textbf{Reliability and Dependability:}
ML models in cybersecurity are particularly fragile with respect to reliability and trustworthiness. Their effectiveness is shaped by dynamic operational environments, adversarial settings, and the quality of training data.

\textbf{Model Drift:}
Cybersecurity applications evolve rapidly, leading to concept and data drift. Concept drift occurs when the target of prediction changes over time, whereas data drift arises when the predictive features shift while the target remains constant. Both are common in cybersecurity, where threat patterns change frequently. As a result, model accuracy deteriorates over time. AutoML pipelines should therefore incorporate continuous monitoring and verification mechanisms. Recent work has begun addressing drift detection and adaptation in AutoML~\cite{Jorge19}.

\textbf{Adversarial and Hostile Settings:}
ML models were not originally designed for adversarial environments~\cite{mcdaniel2016machine}. Yet, in cybersecurity, adversaries routinely craft malicious inputs to mislead models. In AutoML, these risks are amplified: automatically generated models are easier to reverse engineer, clone, and probe compared to handcrafted models. Attackers can exploit probing techniques to reconstruct AutoML models, study their weaknesses, and launch targeted attacks. Pang et al.~\cite{pang2021security} showed that handcrafted deep-learning models are generally more robust against adversarial attacks than AutoML-generated models, underscoring the need to strengthen AutoML in hostile environments.

\textbf{Data Quality:}
Data quality strongly determines ML performance, yet cybersecurity datasets frequently suffer from selection bias, exclusion bias, observer bias, and labeling errors. Imbalanced, noisy, or insufficient training data are common issues~\cite{Abubakar15,Yavanoglu17,Pawlicki20}. Even newly released datasets quickly become outdated due to drift.  

Recent work has proposed automated methods to assess and mitigate data quality issues. DeCastro et al.~\cite{DeCastro21} introduced a multidimensional framework for data quality assessment, leaving remediation decisions to analysts. Other approaches use augmentation or generative models to produce high-quality synthetic data~\cite{Abbiati21, Oesch21}. While we do not expect AutoML tools to generate datasets, they should at least support automated data quality assessment, particularly for frameworks leveraging meta-learning~\cite{Vilalta2002, Dyrmishi19}.

\textbf{Interpretability and Explainability:}
Interpreting ML outputs is essential in cybersecurity, where decisions must be actionable and defensible~\cite{carvalho2019machine, gilpin2018explaining}. Although simple decision trees can be transparent, the complexity of cybersecurity threats often necessitates deep or large trees that are difficult to interpret. Analysts require explanations for why a transaction was flagged as fraud or why a file was classified as malicious in order to design effective security policies and countermeasures.

For AutoML adoption, tools must provide built-in interpretability. If explanations require ML experts, the value of AutoML diminishes. However, recent studies indicate that AutoML pipelines often generate models that are harder to interpret than expert-crafted models~\cite{arzani2021interpret, freitas2019automated}. Enhancing interpretability within AutoML is therefore critical for real-world cybersecurity use.
\section{Methodology}
\label{sec:method}

This section outlines the process used to select the cybersecurity datasets and AutoML tools evaluated in this study.

\subsection{Cybersecurity Dataset Selection}
We initially considered 18 publicly available cybersecurity datasets. The final selection was narrowed to 11 using the following criteria: (i) the dataset must address one of the core cybersecurity domains (malware detection, spam detection, fraud detection, anomaly detection, or intrusion detection); (ii) it must be publicly available; (iii) recognized as a benchmark dataset in the literature; (iv) contain labeled data; and (v) have been evaluated with handcrafted ML methods in the past five years. The final datasets are: CICAndMal2017~\cite{lashkari2018toward}, EMBER2018~\cite{2018arXiv180404637A}, SMS Spam~\cite{almeida2011contributions, SMS_spam}, Nazario Phishing Corpus~\cite{nazario_phishing}, SpamAssassin~\cite{spamassassin}, Enron Email~\cite{enron_email}, UCI Phishing Websites~\cite{uciphishing}, Deceptive Opinion Spam~\cite{ott2011finding}, UNSW-NB15~\cite{moustafa2015comprehensive}, UNSW-Bot-IoT~\cite{koroniotis2019towards}, and CMU Keystroke Dynamics~\cite{killourhy2009comparing}.

Although several AutoML tools advertise data preprocessing, in practice most required tabular input and provided limited preprocessing support. Consequently, we prepared all datasets for compatibility, applying cleaning and preprocessing techniques consistent with prior studies.  

For CICAndMal2017, which contains 10,854 malware and benign samples categorized into Adware, Ransomware, Scareware, and SMS Malware, we followed the approach of Noorbehbahani et al.~\cite{noorbehbahani2019analysis} and focused on ransomware detection. Ten ransomware families were merged and labeled as “1,” while benign samples were labeled as “0.” This preprocessing likely introduced concept and data drift, enabling us to investigate AutoML performance under drift conditions.

\subsection{AutoML Tool Selection}
From an initial pool of 20 AutoML tools, we excluded those no longer maintained, unreliable in operation, or lacking Python support, since language differences could affect runtime comparisons. Ten tools remained: H2O, Neural Network Intelligence (NNI), TPOT, Auto-ViML, AutoKeras, Auto-PyTorch, AutoGluon, FLAML, PyCaret, and Auto-Sklearn. We then refined this list by evaluating automation capabilities for preprocessing and the range of supported models. Auto-PyTorch was excluded due to functional similarity with AutoKeras, while Auto-Sklearn was removed due to compatibility issues. The final selection comprised eight tools: H2O, TPOT, Auto-ViML, AutoKeras, AutoGluon, FLAML, PyCaret, and MLJAR. These tools represent a broad spectrum of automation capabilities, ranging from fully manual to fully automated.

\textbf{Data Preprocessing:}
 Most AutoML tools provide only limited automation, focusing primarily on model selection and hyperparameter optimization. Tools such as Auto-ViML, AutoGluon, and PyCaret perform full data cleaning (e.g., removal of NaN values, variable classification), with PyCaret also allowing manual verification of variable labeling. Auto-ViML, TPOT, PyCaret, and MLJAR provide automated feature selection, while AutoGluon removes duplicate and constant-value features. Feature engineering support varies: H2O includes target encoding and word2vec; TPOT applies functions such as polynomial features and PCA; Auto-ViML integrates Featuretools; AutoGluon generates n-grams, performs binning and encoding; PyCaret offers built-in polynomial, group, and binning functions; and MLJAR automatically creates golden features from feature pairs and applies random feature selection. Overall, data preprocessing remains one of the least mature components of AutoML pipelines.

\textbf{Model Support:}
 Most tools support a wide variety of models, with AutoKeras being the exception due to its focus on neural networks. Auto-ViML also supports fewer models, reflecting its reliance on manual model configuration. With the exception of TPOT, all tools provide at least some NLP support. For image classification, only AutoKeras and AutoGluon include native capabilities within their AutoML pipelines.

\section{Experiment and Discussion}
\label{sec:experiment}

In our setup, we maintained unified hardware and software configurations to compare the performance of ML models generated by AutoML tools with handcrafted models across datasets. This uniformity enabled us to analyze hidden performance patterns, study correlations between tools and datasets, and assess the impact of AutoML compared to expert-built models.

\textbf{Experiment Configuration:}
Experiments were conducted on both Windows and Ubuntu platforms. AutoGluon required Linux, while all other tools were tested on Windows. All experiments were run on the same workstation (12 cores, 3.7 GHz, 32 GB RAM, 1 TB SSD).

Unless predefined splits were available (e.g., EMBER), data were split 75/25 into training and testing sets. Each test was limited to a maximum of two days; incomplete runs were terminated, and interim results recorded. Of the 72 total tests, 70 completed successfully. Recommended or default configurations were used for all tools, and both model performance and runtime were recorded.

\textbf{Evaluation Metrics:}
We used accuracy and balanced accuracy to evaluate performance. Accuracy (Eq.~\ref{eq-accuracy}) summarizes correct predictions over total predictions. While widely reported in prior literature, accuracy alone is insufficient for imbalanced cybersecurity datasets. Balanced accuracy (Eq.~\ref{eq-baccuracy})—the mean of sensitivity and specificity—better reflects performance under class imbalance.

\begin{equation}
\label{eq-accuracy}
accuracy = \frac{(TP + TN)}{TP+FN+FP+TN}
\end{equation}

\begin{equation}
\label{eq-baccuracy}
\begin{split}
sensitivity & = \frac{TP}{(TP+FN)} \\
specificity & = \frac{TN}{(TN+FP)} \\
balanced\_accuracy & = \frac{sensitivity+specificity}{2}\\
\end{split}
\end{equation}

\begin{table*}[h]
\centering
\caption{Balanced accuracy of AutoML tools across datasets}
\label{tab:Average-Balanced-Accuracy}
\resizebox{\textwidth}{!}{
\begin{tabular}{|c|ccccccccc|c|}
\hline
\backslashbox{AutoML Tool}{Dataset}
& UNSW & UCI & Email-1 & SMS & Opinion & Botnet & Enron & CICMal & Ember & \begin{tabular}[c]{@{}c@{}}Average Balanced\\ Accuracy\end{tabular} \\ \hline
Auto-ViML & 0.9827 & 0.9520 & \cellcolor{green!30}0.9244 & \cellcolor{green!30}0.9009 & 0.7965 & \cellcolor{green!30}1.0000 & 0.8969 & 0.6850 & 0.8908 & 0.8921 \\
TPOT & \cellcolor{red!40}0.9803 & 0.9654 & 0.9142 & 0.8933 & 0.8775 & 0.7551 & \cellcolor{green!30}0.9879 & 0.7671 & \cellcolor{red!40}0.8790 & 0.8911 \\
AutoKeras & 0.9838 & 0.9557 & 0.8814 & 0.8685 & 0.8474 & \cellcolor{red!40}0.6783 & & 0.7268 & 0.8839 & 0.8532 \\
H2O & 0.9814 & \cellcolor{red!40}0.9130 & 0.9123 & \cellcolor{red!40}0.7384 & \cellcolor{red!40}0.6006 & 0.7856 & \cellcolor{red!40}0.6684 & 0.5552 & 0.9059 & \cellcolor{red!30}0.7845 \\
AutoGluon & 0.9898 & 0.9662 & \cellcolor{red!40}0.8585 & 0.8531 & 0.8591 & 0.7775 & 0.9405 & \cellcolor{green!30}1.0000 & 0.9248 & 0.9077 \\
FLAML & 0.9905 & 0.9664 & 0.9115 & 0.8823 & 0.8847 & 0.9744 & 0.9458 & \cellcolor{red!40}0.5141 & \cellcolor{green!30}0.9361 & 0.8895 \\
MLJAR & 0.9907 & \cellcolor{green!30}0.9674 & 0.9225 & 0.8867 & 0.8350 & 0.9904 & 0.9543 & 0.7884 & 0.9292 & 0.9183 \\
PyCaret & \cellcolor{green!30}0.9947 & 0.9357 & 0.9111 & 0.8849 & \cellcolor{green!30}0.9747 & 0.9705 & 0.9399 & & 0.9270 & \cellcolor{green!30}0.9423 \\ \hline
\begin{tabular}[c]{@{}c@{}}Average Balanced\\ Accuracy\end{tabular} & 0.9867 & 0.9527 & 0.9045 & 0.8635 & 0.8344 & 0.8665 & 0.9048 & 0.7195 & 0.9096 & \\ \hline
\end{tabular}}
\end{table*}

\begin{table*}[h]
\centering
\caption{Average training time (minutes) by dataset}
\label{tab:Average-Time}
\resizebox{\textwidth}{!}{
\begin{tabular}{|c|ccccccccc|c|}
\hline
& UNSW & UCI & Email-1 & SMS & Opinion & Botnet & Enron & CICMal & Ember & Average Time \\ \hline
Auto-ViML & 108.0288 & 0.4380 & 0.2078 & 0.2275 & 0.9130 & 38.6382 & 203.7065 & 17.4757 & 13.3468 & 42.5536 \\
TPOT & 227.1843 & 12.8118 & 3.9043 & 7.0497 & \cellcolor{red!30}83.8348 & 116.6180 & 281.6133 & \cellcolor{red!30}324.8277 & 343.5528 & \cellcolor{red!30}155.7108 \\
AutoKeras & 183.4265 & 6.3808 & 0.9142 & 2.3072 & 39.5948 & \cellcolor{red!30}296.0975 & & 79.5768 & \cellcolor{red!30}448.4963 & 132.0993 \\
H2O & \cellcolor{green!30}52.9688 & \cellcolor{red!30}40.5953 & \cellcolor{red!30}40.5687 & \cellcolor{red!30}40.5583 & 37.7577 & 69.3332 & 43.6455 & 52.6193 & 58.6183 & 48.5184 \\
AutoGluon & 105.0398 & \cellcolor{green!30}0.3505 & 0.1763 & \cellcolor{green!30}0.2138 & \cellcolor{green!30}0.4485 & \cellcolor{green!30}21.3143 & \cellcolor{green!30}16.7840 & \cellcolor{green!30}5.6268 & \cellcolor{green!30}10.0090 & \cellcolor{green!30}17.7737 \\
FLAML & 32.9498 & 12.8398 & 2.8675 & 27.6833 & 32.7937 & 45.1928 & 18.0927 & 50.2392 & 59.6468 & 31.3673 \\
MLJAR & 60.5492 & 23.7375 & 7.0333 & 9.2987 & 60.1502 & 59.8412 & 70.0713 & 59.4757 & 56.3665 & 45.1693 \\
PyCaret & \cellcolor{red!30}243.1398 & 1.0002 & \cellcolor{green!30}0.1208 & 0.7495 & 4.8793 & 69.4185 & \cellcolor{red!30}668.8400 & & 152.2065 & 142.5443 \\ \hline
Average Time & 126.6609 & 12.2692 & 6.9741 & 11.0110 & 32.5465 & 89.5567 & 186.1076 & 84.2630 & 142.7804 & \\ \hline
\end{tabular}}
\end{table*}

\begin{table*}[h]
\centering
\caption{Average accuracy of AutoML tools across datasets}
\label{tab:Average Accuracy}
\resizebox{\textwidth}{!}{
\begin{tabular}{|c|ccccccccc|c|}
\hline
\backslashbox{AutoML Tool}{Dateset}
& UNSW & UCI & Email-1 & SMS & Opinion & Botnet & Enron & CICMal & Ember & Average Accuracy \\ \hline
Auto-ViML & \cellcolor{red!30}0.9895 & 0.9518 & \cellcolor{green!30}0.9693 & 0.9605 & 0.7958 & \cellcolor{green!30}1.0000 & 0.9443 & 0.6807 & 0.9244 & 0.9129 \\
TPOT & 0.9930 & 0.9674 & 0.9553 & 0.9641 & 0.8775 & 0.9982 & \cellcolor{green!30}0.9926 & 0.7671 & 0.8761 & 0.9324 \\
AutoKeras & 0.9924 & 0.9573 & 0.9448 & 0.9555 & 0.7884 & 0.9639 & & 0.7279 & 0.8812 & 0.9014 \\
H2O & 0.9905 & \cellcolor{red!30}0.9150 & 0.9619 & \cellcolor{red!30}0.9189 & \cellcolor{red!30}0.6100 & \cellcolor{red!30}0.9531 & \cellcolor{red!30}0.8747 & 0.5591 & \cellcolor{red!30}0.9034 & \cellcolor{red!30}0.8541 \\
AutoGluon & 0.9956 & 0.9689 & 0.9264 & 0.9555 & 0.8600 & 0.9999 & 0.9777 & \cellcolor{green!30}1.0000 & 0.9218 & \cellcolor{green!30}0.9562 \\
FLAML & 0.9960 & 0.9689 & 0.9606 & 0.9641 & \cellcolor{green!30}0.8850 & 0.9999 & 0.9824 & \cellcolor{red!30}0.5094 & \cellcolor{green!30}0.9339 & 0.9111 \\
MLJAR & \cellcolor{green!30}0.9961 & \cellcolor{green!30}0.9696 & 0.9593 & 0.9641 & 0.8350 & 0.9997 & 0.9749 & 0.7882 & 0.9267 & 0.9348 \\
PyCaret & 0.9917 & 0.9370 & \cellcolor{red!30}0.9212 & \cellcolor{green!30}0.9648 & 0.8750 & 0.9298 & 0.9768 & & 0.9290 & 0.9407 \\ \hline
Average Accuracy & 0.9924 & 0.9538 & 0.9500 & 0.9560 & 0.8108 & 0.9782 & 0.9544 & 0.6996 & 0.8577 & \\ \hline
\end{tabular}}
\end{table*}

\subsection{Experiment Results}
As seen in Table \ref{tab:Average-Balanced-Accuracy}, all results are present except for AutoKeras with the Enron dataset and PyCaret with the CICMal dataset. These configurations were unable to return results within the 2-day test period. Table \ref{tab:Average-Balanced-Accuracy} shows balanced accuracy values; Table \ref{tab:Average-Time} shows training time in minutes; Table \ref{tab:Average Accuracy} shows accuracy.

\begin{figure*}[h!]
  \includegraphics[width=\linewidth]{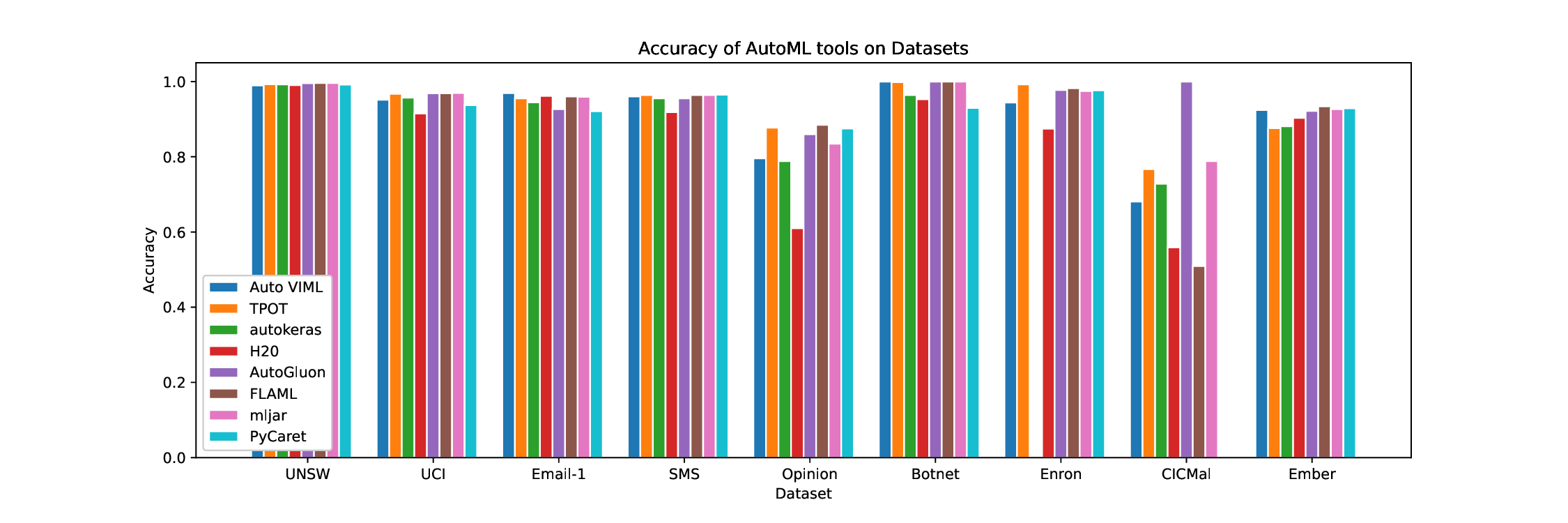}
  \caption{Balanced accuracy of AutoML tools across datasets}
  \label{fig:all_baccuracy}
\end{figure*}

Comparing Tables \ref{tab:Average-Balanced-Accuracy} and \ref{tab:Average Accuracy}, the ranking of tools differs by metric. AutoGluon achieved the highest average accuracy (95.66\%), followed by PyCaret (94.07\%). Using balanced accuracy, PyCaret ranked first (94.23\%) and AutoGluon third (90.77\%). Balanced accuracy is the more suitable metric under class imbalance.

\textbf{AutoML Tools Performance:}
%Comparing all of the AutoML tools across all the testeddatasets, we looked at two metrics, balanced accuracy and time. At first glance, it seems unattainable to determine the best overall tool based on these two metrics. The inconsistent performance of the tools on the different datasets  (as seen in Fig.\ref{fig:all_baccuracy}) and some tools' inability to complete all tests made it difficult to agree on an overall best tool. 
%For instance, PyCaret is ranked as the best tool based on all evaluated tools' reported average balanced accuracy. However, PyCaret failed to generate a ML model for the CICMal dataset within the allowed time. 
%If we consider this failure equal to zero balanced accuracies, then PyCaret will rank last. However, if PyCaret was given more time, it could generate a ML model for CICMal.

Fig. \ref{fig:all_baccuracy} presents a comparison of balanced accuracy across all AutoML tools across tested datasets. Initially, determining the best overall tool based on these metrics seems challenging due to inconsistent performance and some tools' inability to complete all tests. MLJAR is considered the best AutoML tool based on the observed average balanced accuracy as it ranked \#1 on one dataset (UCI) and \#2 on most other datasets.  Furthermore, it successfully finished all the tests within the designated time frame.

%If we need to agree on one best AutoML tool based on the observed average balanced accuracy, we believe it is MLJAR. While MLJAR placed first on only a  single dataset (UCI), it was ranked second on nearly all the other datasets.  In addition, it completed all the tests within the allowed time.

When evaluating AutoML tools, the time to generate an ML model is crucial, especially in time-sensitive domains where we train and update the models continuously.  compares AutoML tool training times across multiple datasets. 

When assessing AutoML tools, the time required to generate a machine learning model is critical, especially in time-sensitive domains where continuous training and updating of the ML models is anticipated. In Fig.~\ref{fig:avg_time} compares the training times of AutoML tools for several datasets. Details on timing values can be found in Table \ref{tab:Average-Time}.

%The time needed to generate a machine learning model is essential when evaluating AutoML tools, particularly in time-sensitive domains where we expect to train and update the ML models continually. Fig.~\ref{fig:avg_time} compares the training times of AutoML tools for several datasets. The exact timing values are presented in Table \ref{tab:Average-Time} for further detailed analysis.

The results in Fig.~\ref{fig:avg_time} and Table \ref{tab:Average-Time} show no correlation between the time an AutoML tool took to generate a ML model and the models' performance. 
%In other words, we could not anticipate the performance of the generated ML model based on the time an AutoML took to build the model. 
For instance, PyCaret reported the best accuracy and the longest time to generate a ML model for the UNSW dataset. Nevertheless, for every other dataset, the best accuracy did not occur with the longest time to generate the model. Moreover, in some cases, the worst accuracy occurred with the longest or second longest time to generate the model. This is obvious for H2O with the UCI dataset, AuoKeras with the Botnet dataset, and TPOT with the Ember dataset. 
In addition, in several cases for the same dataset, the variation in time between two AutoML tools was enormous, while the variation in their performance was almost negligible. For instance, TPOT reported an accuracy of 98\% on the Enron dataset at quadruple the time MLJAR took to generate a model with 95\% accuracy (see Tables \ref{tab:Average-Balanced-Accuracy} and \ref{tab:Average-Time}). 

The AutoML tools in our study each support various ML models. Most AutoML tools predominantly utilize tree-based and ensemble ML models, as indicated in Tables \ref{tab:BestModels}, \ref{tab:Best-Model}, and \ref{tab:Best-Model-Balanced}. This suggests that these ML models may be more appropriate for data-driven cybersecurity issues or that they reach convergence more quickly than alternative models.

\textcolor{black}{
We observed that some datasets did not meaningfully challenge the AutoML tools. For example, all AutoML tools achieved outstanding balanced accuracy (above 90\%, and on average over 95\%) on the UNSW and UCI datasets. This suggests that security analysts should exercise caution when using datasets where AutoML tools or ML models achieve near-perfect performance with little effort, as such datasets may not reflect the complexity of real-world scenarios. 
}
\textcolor{black}{
For additional insights, we applied the Friedman test to assess whether differences in balanced accuracy across AutoML tools were statistically significant. The test revealed significant variation ($\chi^2$ = 17.70, p = 0.0134). We then conducted Nemenyi post-hoc tests, summarized in Table~\ref{tab:Nemenyi}, which showed that MLJAR significantly outperformed AutoKeras (p = 0.032) and H2O (p = 0.012). No other pairwise differences reached significance. These findings support the conclusion that the choice of AutoML tool can have a meaningful impact on cybersecurity model performance.
}
\begin{table*}[h]
\centering
\caption{Nemenyi Post-hoc p-values for AutoML Tool Pairwise Comparisons (Significant p-values highlighted)}
\label{tab:Nemenyi}
\resizebox{\textwidth}{!}{
\begin{tabular}{|l|c|c|c|c|c|c|c|c|}
\hline
 & AutoVIML & TPOT & AutoKeras & H2O & AutoGluon & FLAML & MLJAR & PyCaret \\
\hline
AutoVIML   & 1.000 & 1.000 & 0.786 & 0.601 & 1.000 & 0.995 & 0.729 & 1.000 \\
TPOT       & 1.000 & 1.000 & 0.786 & 0.601 & 1.000 & 0.995 & 0.729 & 1.000 \\
AutoKeras  & 0.786 & 0.786 & 1.000 & 1.000 & 0.881 & 0.288 & \cellcolor{red!30}0.032 & 0.786 \\
H2O        & 0.601 & 0.601 & 1.000 & 1.000 & 0.786 & 0.156 & \cellcolor{red!30}0.012 & 0.601 \\
AutoGluon  & 1.000 & 1.000 & 0.881 & 0.786 & 1.000 & 0.980 & 0.601 & 1.000 \\
FLAML      & 0.995 & 0.995 & 0.288 & 0.156 & 0.980 & 1.000 & 0.989 & 0.995 \\
MLJAR      & 0.729 & 0.729 & \cellcolor{red!30}0.032 & \cellcolor{red!30}0.012 & 0.601 & 0.989 & 1.000 & 0.729 \\
PyCaret    & 1.000 & 1.000 & 0.786 & 0.601 & 1.000 & 0.995 & 0.729 & 1.000 \\
\hline
\end{tabular}
}
\end{table*}

\textbf{Data-Driven Cybersecurity Problems with AutoML:}
%Our study covered datasets from different cybersecurity domains, including malware detection, network intrusion, phishing, opinion spam, and biometric authentication. The AutoML tools generally showed acceptable performance across all of these domains. When looking at the performance of AutoML across different cybersecurity domains, we see two groups (as seen in Fig. \ref{fig:bavg_std} and \ref{fig:std}). The first group comprises the datasets where all the AutoML tools reported outstanding or comparable accuracy. %The second group contains the datasets where most of the AutoML tools did not succeed in reporting good accuracy results, and the results showed noticeable deviations between the tools.
%The datasets comprising the second group exhibit significant discrepancies in accuracy estimates across the AutoML tools and failed to produce optimal results for the majority of them. We will focus our discussion on the second group, which includes five datasets: Opinion Spam, CICMalware, Enron, UNSW Botnet, and the CMU Keystroke dataset.

\begin{figure}[!h]
\centering
  \includegraphics[scale=0.4]{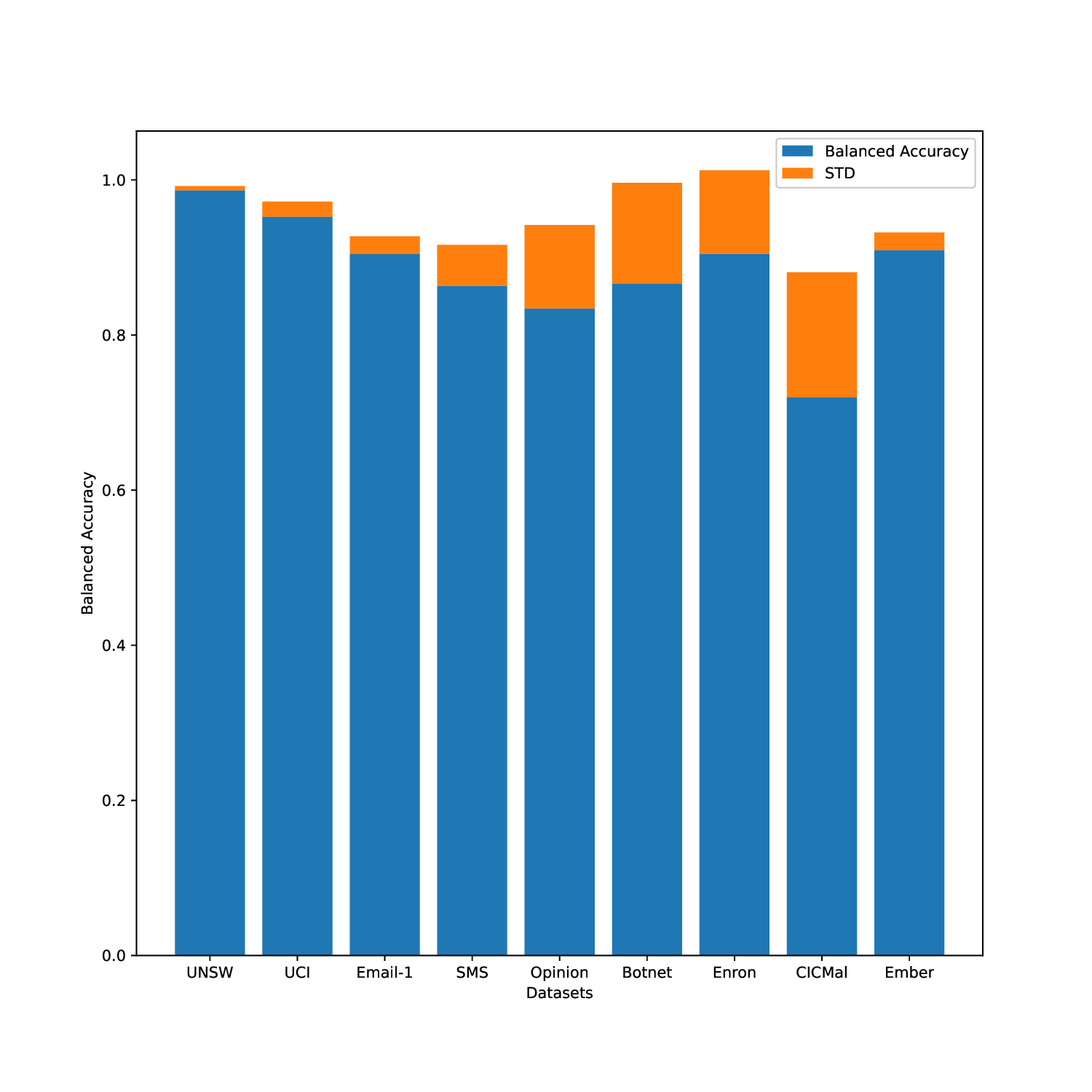}
  \caption{Average balanced accuracy and corresponding standard deviation reported by AutoML tools for each dataset.}
  \label{fig:bavg_std}
\end{figure}

Our study analyzed datasets from various cybersecurity domains, including malware detection, network intrusion, phishing, opinion spam, and biometric authentication. When assessing AutoML performance across these domains, we identified two groups (refer to Fig. \ref{fig:bavg_std}). The first group consists of datasets where all AutoML tools achieved outstanding or comparable accuracy. Conversely, the second group exhibited significant accuracy variations among AutoML tools, resulting in suboptimal outcomes for most. Our discussion will center on this second group, comprising five datasets: Opinion Spam, CICMalware, Enron, UNSW Botnet, and the CMU Keystroke dataset.

%\textbf{Opinion Spam} had the best result of 97.47\%  (balanced accuracy) achieved by PyCaret, and the second best result was 88.47\%, achieved by MLJAR, as shown in Table \ref{tab:Average-Balanced-Accuracy} . The best result was achieved using the logistic regression ML model, while all the other AutoML tools' results were achieved by tree-based ML models, as shown in Table \ref{tab:BestModels}

PyCaret dominated \textbf{Opinion Spam} detection with a balanced accuracy of 97.47\% (Table \ref{tab:Average-Balanced-Accuracy}), surpassing MLJAR's 88.47\%. Notably, PyCaret achieved this using a logistic regression model, unlike other AutoML tools that relied on tree-based models (Table \ref{tab:BestModels}).

%\textcolor{black}{\textbf{Botnet} had the best result achieved by Auto VIML using XGboost ML model. However, deviation in the results is clear as four tools reported balanced accuracy above 97\%, and the other four tools reported balance accuracy less than 79\%, as shown in Table \ref{tab:Average-Balanced-Accuracy}. It is known that XGboost and other tree-based ML models tend to overfit the data, mainly if the trees are too deep with noisy data. This could explain the exceptional results reported by  Auto VIML (100\% accuracy).}

Auto VIML's XGBoost model achieved 100\% accuracy for \textbf{Botnet detection} (Table \ref{tab:Average-Balanced-Accuracy}), but results varied significantly across tools. While four tools achieved balanced accuracy over 97\%, the rest fell below 79\%. Tree-based models like XGBoost tend to overfit when the trees are too deep with noisy data, potentially explaining Auto VIML's exceptional result.

%\textbf{Enron} had a best result of 98.79\% achieved by TPOT, using a GaussianNB ML model, as shown in Tables \ref{tab:Average-Balanced-Accuracy} and \ref{tab:BestModels}. The second-best result was 95.53\% achieved by MLJAR using a LightGBM ML model (tree-based). All tree-based ML models reported an average balanced accuracy of almost 94\% and significantly better runtime than the GaussianNB with TPOT. As shown in Table \ref{tab:Average-Time}, MLJAR using a LightGBM was four times faster than TPOT with GaussianNB. However, neural networks performed poorly on this dataset, failed to generate any ML model for Enron, and H2O using neural network reported the worst accuracy of 66.84\%

\textbf{Enron} demonstrated a peak performance of 98.79\% achieved by TPOT utilizing a GaussianNB ML model, as depicted in Tables \ref{tab:Average-Balanced-Accuracy} and \ref{tab:BestModels}. The second-best result was 95.53\%, attained by MLJAR employing a LightGBM ML model (tree-based). All tree-based ML models showcased an average balanced accuracy nearing 94\%, along with significantly enhanced runtime compared to GaussianNB with TPOT. As illustrated in Table \ref{tab:Average-Time}, MLJAR utilizing LightGBM was four times faster than TPOT with GaussianNB. Nevertheless, neural networks exhibited poor performance on this dataset, failing to produce any ML model for Enron. H2O using neural networks reported the lowest accuracy of 66.84\%.

\textcolor{black}{\textbf{CICMal}, as previously mentioned, had a model drift effect introduced when we assigned the same label to all ransomware families and prepared the test data to contain malicious samples that were not present in the training data. Therefore, we expected that the AutoML tools would report underperforming results. This was true for all the tools except AutoGluon, which reported a perfect accuracy of 100\% using a random forest ML model, as shown in Table \ref{tab:Average-Balanced-Accuracy}. AutoGluon reported perfect accuracy and extremely impressive runtime compared to other tools. AutoGluon was 12 times faster than the second-best tool (MLJAR, with an accuracy of 78.84\%) and 3.5 times faster than any other tool, as shown in Table \ref{tab:Average-Time}.}
\textcolor{black}{We uncovered the mystery of AutoGluon performance with an in-depth analysis of the AutoGluon on CICMal dataset. 
%One of the features of the CICMal dataset is a string representing a time when the network flow containing a binary (benign or malicious) has occurred. While all AutoML tools should ignore this time feature (because it is considered a random string), AutoGluon can recognize strings in well-known formats and convert them to numerical values, such as timestamps. 
The CICMal dataset includes a timestamp feature, which is a string denoting the moment a network event occurred. While it is desirable for any AutoML tool to disregard the timestamp due to it being regarded as a random string, AutoGluon identified it as a feature and performed a numerical conversion.
%All the malware binaries in the CICMal dataset were injected at unique timestamps. 
Consequently, the timestamp feature emerged as an exceedingly dominant feature and a tree-based algorithm achieved almost flawless accuracy. 
%It is important to note that the timestamp of the malware occurrence is a falsely dominant feature. 
Therefore, we removed the timestamp feature prior to employing AutoGluon on the CICMal dataset, resulting in a decrease in accuracy to 86.94\%. Additionally, the ML model switched from a random forest to WeightedEnsemble ML model. 
%The runtime remained the same, but we are curious if AutoGloun found another shortcut with its feature engineering built-in functions.
}

\textbf{CMU Keystroke Dynamics} represents a One-Class Classification (OCC) problem~\cite{occ18}. Many cybersecurity tasks naturally fall into this category, including zero-day vulnerability detection, biometric authentication, and fraud detection. Designing effective ML models for OCC remains challenging, and current AutoML tools provide limited support. In our study, only H2O and PyCaret offered OCC capabilities, achieving 74.7\% balanced accuracy (72.6\% overall accuracy) and 86.2\% balanced accuracy (85.1\% overall accuracy), respectively. By contrast, handcrafted models applied to the CMU Keystroke dataset have reported accuracies exceeding 95\%~\cite{cmuKS17, cmuKS19}, highlighting a significant performance gap and the need for enhanced AutoML support in OCC tasks.

%, where PayCart supports different ML models that support OCC or anomaly detection

%For OCC, handcrafted ML models will generally outperform AutoML, and several handcrafted models applied to the CMU Keystroke datasets reported accuracy above 95\% \cite{cmuKS17, cmuKS19}.}

\begin{table*}[]
\centering
\caption{Models Used by Each AutoML Tool}
\label{tab:BestModels}
\resizebox{\textwidth}{!}{
\begin{tabular}{|c|ccccccc|}
\hline
        & Auto VIML     & TPOT              & H20               & AutoGluon         & FLAML             & MLJAR    & PyCaret                  \\ \hline
UNSW    & XGBoost       & Decision Tree     & Gradient Boosting & WeightedEnsemble  & Gradient Boosting & LightGBM & XGBoost                  \\
UCI     & XGBoost       & XGBoost           & Stacked Ensembles & Neural Networks   & XGBoost           & Ensemble & ExtraTrees               \\
Email-1 & Random Forest & XGBoost           & Stacked Ensembles & CatBoost          & XGBoost           & Ensemble & Random Forest            \\
SMS     & Random Forest & Random Forest     & Neural Networks   & Neural Networks   & Random Forest     & Ensemble & CatBoost                 \\
\cellcolor{blue!30}Opinion & \cellcolor{red!30}Random Forest & \cellcolor{red!30}Gradient Boosting & \cellcolor{red!30}Stacked Ensembles & \cellcolor{red!30}ExtraTrees        & \cellcolor{red!30}Gradient Boosting & \cellcolor{red!30}CatBoost & \cellcolor{green!30}Logistic Regression      \\
Botnet  & XGBoost       & DecisionTree      & Gradient Boosting & Random Forest     & ExtraTrees        & Ensemble & Random Forest Classifier \\
Enron   & Random Forest & GaussianNB        & Neural Networks   & WeightedEnsemble  & Gradient Boosting & LightGBM & Naive Bayes              \\
\cellcolor{blue!30}CICMal  & XGBoost       & Random Forest     & Stacked Ensembles & \cellcolor{red!30}Random Forest     & ExtraTrees        & Ensemble &                          \\
\cellcolor{blue!30}Ember   & XGBoost       & \cellcolor{red!30}Gradient Boosting & Stacked Ensembles & Gradient Boosting & \cellcolor{green!30}Gradient Boosting & Ensemble & XGBoost                  \\ \hline
\end{tabular}}
\end{table*}

\begin{table}[!h]
\centering
\caption{Best Model for Each Dataset (Average Accuracy)}
\label{tab:Best-Model}
\begin{tabular}{|cccc|}
\hline
Dataset & Best Tool & Model Used        & Accuracy \\ \hline
UNSW    & MLJAR     & LightGBM          & 0.9961   \\
UCI     & MLJAR     & Ensemble          & 0.9696   \\
Email-1 & Auto VIML & Random Forest     & 0.9693   \\
SMS     & PyCaret   & CatBoost          & 0.9648   \\
Opinion & FLAML     & Gradient Boosting & 0.8850   \\
Botnet  & Auto VIML & XGBoost           & 1.0000   \\
Enron   & TPOT      & GaussianNB        & 0.9926   \\
CICMal  & AutoGluon & Random Forest     & 1.0000   \\
Ember   & FLAML     & Gradient Boosting & 0.9339   \\ \hline
\end{tabular}
\end{table}

\begin{table}[!h]
\centering
\caption{Best Model for Each Dataset (Balanced Accuracy)}
\label{tab:Best-Model-Balanced}
\begin{tabular}{|cccc|}
\hline
Dataset & Best Tool & Model Used          & Balanced Accuracy \\ \hline
UNSW    & PyCaret   & XGBoost             & 0.9947            \\
UCI     & MLJAR     & Ensemble            & 0.9674            \\
Email-1 & Auto VIML & Random Forest       & 0.9244            \\
SMS     & Auto VIML & Random Forest       & 0.9009            \\
Opinion & PyCaret   & Logistic Regression & 0.9747            \\
Botnet  & Auto VIML & XGBoost             & 1.0000            \\
Enron   & TPOT      & GaussianNB          & 0.9879            \\
CICMal  & AutoGluon & Random Forest       & 1.0000            \\
Ember   & FLAML     & Gradient Boosting   & 0.9361            \\ \hline
\end{tabular}
\end{table}

\begin{figure*}[!h]
  \includegraphics[width=\linewidth]{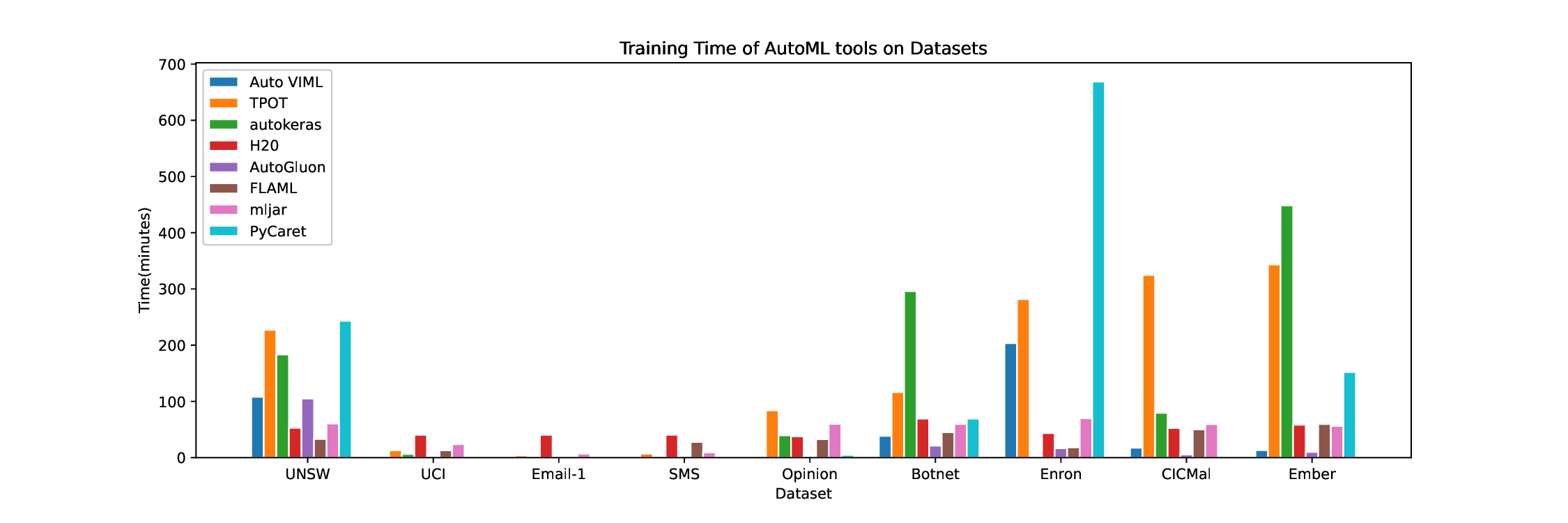}
  \caption{Comparison of AutoML tools for several datasets with respect to model training time.}
  \label{fig:avg_time}
\end{figure*}

\subsection{AutoML Best Practices}
AutoML is a powerful technology for enabling ML on a large scale. However, the promise that AutoML will simplify the construction of ML pipelines is only partially true. Our study clearly shows that selecting the best-suited AutoML is not trivial. ML practitioners are used to compare and evaluate many ML models to find the best one for a given problem. This is also the case with AutoML. The developers must compare and evaluate several AutoML solutions before finding the one that fits their needs.

While AutoML streamlines hyperparameter tuning and model selection, it is essential to understand how different AutoML tools work and can be configured. Using any AutoML tool without understanding how it works is dangerous, as it can quickly generate a biased model. For instance, when AutoML generates a tree-based ML model, it is essential to consider how easily a tree-based algorithm could overfit the data. 

Ensemble models generated by  AutoML are complex to interpret and debug; we could not identify the underlying ML models used to build the ensemble model in several cases. We agree with Xue et al.\cite{xue2019transferable} that using meta-learning could improve the performance of AutoML tools and reduce the time needed for model selection and hyperparameter tuning. In addition, it is a good idea for complex domains such as cybersecurity to integrate a feature store \cite{featstore21} with predefined patterns for feature engineering and extraction.

\begin{table*}[]
\centering
\caption{Models used by each AutoML tool across datasets}
\label{tab:BestModels}
\resizebox{\textwidth}{!}{
\begin{tabular}{|c|ccccccc|}
\hline
 & Auto-ViML & TPOT & H2O & AutoGluon & FLAML & MLJAR & PyCaret \\ \hline
UNSW & XGBoost & Decision Tree & Gradient Boosting & WeightedEnsemble & Gradient Boosting & LightGBM & XGBoost \\
UCI & XGBoost & XGBoost & Stacked Ensembles & Neural Networks & XGBoost & Ensemble & ExtraTrees \\
Email-1 & Random Forest & XGBoost & Stacked Ensembles & CatBoost & XGBoost & Ensemble & Random Forest \\
SMS & Random Forest & Random Forest & Neural Networks & Neural Networks & Random Forest & Ensemble & CatBoost \\
\cellcolor{blue!30}Opinion & \cellcolor{red!30}Random Forest & \cellcolor{red!30}Gradient Boosting & \cellcolor{red!30}Stacked Ensembles & \cellcolor{red!30}ExtraTrees & \cellcolor{red!30}Gradient Boosting & \cellcolor{red!30}CatBoost & \cellcolor{green!30}Logistic Regression \\
Botnet & XGBoost & Decision Tree & Gradient Boosting & Random Forest & ExtraTrees & Ensemble & Random Forest Classifier \\
Enron & Random Forest & GaussianNB & Neural Networks & WeightedEnsemble & Gradient Boosting & LightGBM & Naïve Bayes \\
\cellcolor{blue!30}CICMal & XGBoost & Random Forest & Stacked Ensembles & \cellcolor{red!30}Random Forest & ExtraTrees & Ensemble & \\
\cellcolor{blue!30}Ember & XGBoost & \cellcolor{red!30}Gradient Boosting & Stacked Ensembles & Gradient Boosting & \cellcolor{green!30}Gradient Boosting & Ensemble & XGBoost \\ \hline
\end{tabular}}
\end{table*}

\begin{figure*}[!h]
  \includegraphics[width=\linewidth]{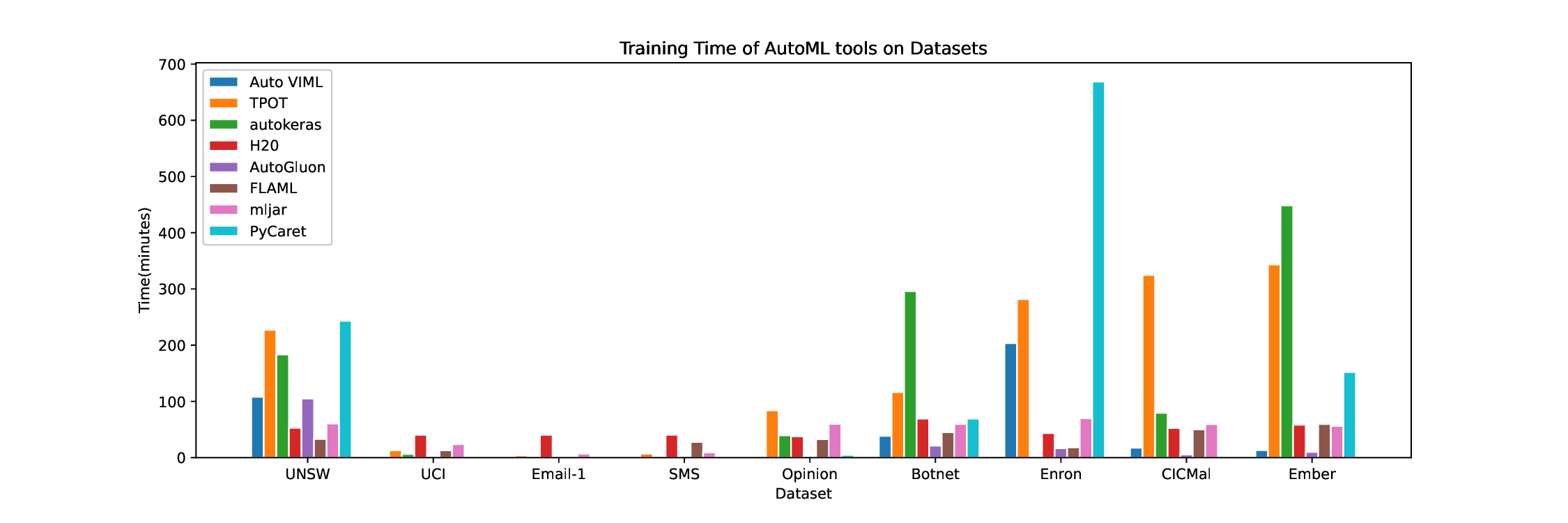}
  \caption{Training time of AutoML tools across datasets}
  \label{fig:avg_time}
\end{figure*}

\section{Conclusion}
\label{sec:conclusion}

This study presented a systematic evaluation of eight AutoML frameworks across eleven publicly available cybersecurity datasets, covering diverse tasks such as intrusion detection, malware classification, phishing, fraud detection, and spam filtering. Our results demonstrate that while AutoML tools provide a valuable degree of automation, their performance varies substantially across datasets and no single solution consistently outperforms others. This highlights a paradigm shift in machine learning workflows: the challenge is less about choosing an individual algorithm and more about selecting and configuring the most suitable AutoML framework.

We observed that AutoML systems often favor tree-based and ensemble methods, which generally perform well but raise concerns of overfitting and limited interpretability. Moreover, runtime efficiency showed no clear correlation with predictive performance, underscoring the need to balance both factors when applying AutoML in time-sensitive cybersecurity environments. Critical challenges remain, particularly around coping with model drift, defending against adversarial manipulation, and improving the interpretability of AutoML-generated models.

Future work should explore meta-learning and automated feature engineering to enhance efficiency and responsiveness, as well as investigate mechanisms for continuous adaptation to evolving threats and streaming data. Equally important is the integration of domain expertise into AutoML workflows, enabling collaboration between human analysts and automated systems. Addressing these directions will be essential for advancing AutoML toward trustworthy, robust, and practical deployment in high-stakes cybersecurity contexts.

\section*{Disclosure Statement}
This manuscript was prepared with the assistance of AI-based language editing tools (ChatGPT by OpenAI) for grammar correction, text tightening, and improvement of academic tone. The authors reviewed and approved all AI-assisted edits to ensure accuracy and accountability for the final content.

\bibliographystyle{elsarticle-num}
\bibliography{mybibliography}

\end{document}